\documentclass[final]{svjour2}
\usepackage{graphicx}
\usepackage{rotating}
\usepackage{amssymb}
\usepackage{amsmath}
\usepackage{bm}
\usepackage[square,numbers]{natbib}
\makeatletter
\journalname{Journal of Low Temperature Physics}
\begin{document}

\newcommand{\hdblarrow}{H\makebox[0.9ex][l]{$\downdownarrows$}-}
\title{Relaxation of Bose-Einstein Condensates of Magnons in
  Magneto-Textural Traps in Superfluid $^3$He-B}

\author{P.J. Heikkinen \and S. Autti \and V.B. Eltsov \and J.J.~Hosio
\and M. Krusius \and V.V. Zavjalov}

\institute{
P.J. Heikkinen \and S. Autti \and V.B. Eltsov \and J.J.~Hosio \and M. Krusius \and V.V. Zavjalov \at
O.V. Lounasmaa Laboratory, Aalto University, P.O. Box 15100, 00076 AALTO, Finland\\
%Tel.: +358 40 727 6892\\
% Fax:\\
\email{petri.heikkinen@aalto.fi}
}
\date{Received: date / Accepted: date}

\maketitle

\begin{abstract}

In superfluid $^3$He-B externally pumped quantized spin-wave excitations or magnons spontaneously form a
  Bose-Einstein condensate in a 3-dimensional trap created with the
  order-parameter texture and a shallow minimum in the polarizing field. The condensation
  is manifested by coherent precession of the magnetization with a common
  frequency in a large volume. The trap
  shape is controlled by the profile of the applied magnetic
  field and by the condensate itself via
  the spin-orbit interaction. The trapping potential can be experimentally
  determined with the spectroscopy of the magnon levels in the trap. We
  have measured the decay of the ground state condensates
  after switching off the pumping in the temperature range $(0.14\div
  0.2)T_{\mathrm{c}}$. Two contributions to the relaxation
  are identified: (1) spin diffusion with the diffusion coefficient
  proportional to the density of thermal quasiparticles and (2) the approximately temperature-independent radiation
  damping caused by the losses in the NMR pick-up circuit. The measured dependence of
  the relaxation on the shape of the trapping potential is in a good agreement
  with our calculations based on the
  magnetic field profile and the magnon-modified texture. Our values for the spin
  diffusion coefficient at low temperatures agree with the
  theoretical prediction and earlier
  measurements at temperatures above $0.5T_{\mathrm{c}}$.


\end{abstract}

\section{Introduction}

One of the interesting recent developments in condensed-matter physics
is the extension of the concept of Bose-Einstein condensation to bosonic
quasiparticles in solid-state and liquid systems~\cite{snoke_nature,bunvol_review}. This concept is
applicable to systems of pumped quasiparticles if the partial equilibrium within such a system with an approximately constant number of quasiparticles is reached faster than the full thermodynamic equilibrium, where the quasiparticle number is
not conserved. By pumping, the quasiparticle number can be kept close to constant
and the system stays in a dynamic steady state. Alternatively, one can study slowly
decaying condensates after the pumping is stopped, which is similar
to the case of a condensate of cold atoms in an optical trap where the particle number continuously
decreases due to evaporation and recombination processes. Bose-Einstein condensation
of quasiparticles has been demonstrated for magnons in various systems~\cite{bunvol_prl98,demokritov_nature,bunkov_kazan},
for exciton-polaritons~\cite{kasprzak_nature}, and even for photons~\cite{klaers_nature}.

In superfluid $^3$He-B magnons are pumped using NMR techniques and their
condensation is manifested by spontaneous long-lived coherent precession of the magnetization in an external magnetic field. At temperatures above 0.4$T_{\mathrm{c}}$
the condensate emerges as a so called homogeneously precessing domain
(HPD)~\cite{borovik_HPD,fomin_HPD}. In the HPD the uniformly precessing magnetization is deflected from the magnetic field by an angle exceeding $104^\circ$. The domain typically occupies a large fraction of the sample volume bounded by the walls of the container. If part of the sample is not involved in
the precession, it is separated from the HPD by a relatively thin
boundary where the tipping angle of magnetization rapidly changes. In the
absence of pumping the HPD decays owing to the Leggett-Takagi relaxation
originating from the coupling of spins of the normal and superfluid
components, and owing to the spin diffusion through the boundary~\cite{Bunkov_etal_PRL65}.
Additionally, at temperatures below about
0.4$T_{\mathrm{c}}$ the HPD experiences highly accelerated decay known as
``catastrophic relaxation" owing to the Suhl type instability~\cite{bunkov_suhl_EL}.

At even lower temperatures, below about $0.3T_{\mathrm{c}}$, another type of
magnon BEC, the so-called trapped condensate, is observed in
$^3$He-B~\cite{lancaster_1st_prl,bunvol_prl98}. In this case magnons are confined to a 3-dimensional trap formed
by the combined effect of the spin-orbit interaction in a spatially
inhomogeneous texture of the order-parameter orbital anisotropy axis and the minimum of the Zeeman energy
in the applied magnetic field, Fig.~\ref{fig:meas_setup}. Such traps can be prepared within bulk $^3$He-B
far from walls. The magnon density in such
condensates can be much smaller than in the HPD resulting in tipping angles of the
magnetization $\beta_M \sim 1^\circ \div 10^\circ$ or less. Also there is
no sharp boundary of the condensate, but $\beta_M$ decreases gradually from the center of the
trap as prescribed by an appropriate eigenstate wave function in the trap.

At temperatures where the trapped magnon condensates are observed, Leggett-Takagi relaxation becomes negligible and spin diffusion was
identified as the main mechanism responsible for the temperature dependence
of the relaxation of such condensates~\cite{relax_lancaster}. As the density of the normal
component rapidly decreases with decreasing temperature, the lifetime of
the condensates increases and can exceed half an
hour in the case when the condensate is well isolated from the
walls~\cite{lancaster_long_decay}. Eventually, the lifetime is limited by an approximately
temperature-independent relaxation process~\cite{relax_lancaster} which has not been identified before
this work.

One of the characteristic features of the condensates in the
magneto-textural trap comes from the fact that the order-parameter texture
is not perfectly rigid and can be modified in the presence of the
precessing spin owing to the spin-orbit interaction, i.e., magnons affect
their trapping potential. This ``self-trapping'' property has many
interesting consequences~\cite{bunkov_JLTP138,magnon_PRL}. In particular, it explains why the frequency of
the precession in the condensate increases during relaxation. However,
simultaneously this makes the analysis of the relaxation processes more
complicated. In this report we present measurements on the relaxation of
the magnon condensates at relatively low density, $\beta_M \lesssim
5^\circ$, where the back reaction on the texture is small. In this case the
relaxation of the transverse magnetization $M_\perp$ is perfectly exponential
$M_\perp \propto \exp(-t/\tau_M)$.

The temperature dependence of the relaxation time constant $\tau_M$ confirms the contributions from spin diffusion together with a
temperature-independent relaxation mechanism.
With the spectroscopy of the
ground and excited magnon levels in the trap we determine accurately the
trapping potential and the corresponding condensate wave functions,
which allows us to calculate values of the spin diffusion coefficient from
the measured $\tau_M$. The results are in a good agreement with
D.~Einzel's theory~\cite{Einzel_JLTP84} which earlier was favorably compared with
measurements of spin diffusion at higher temperatures using HPD~\cite{Bunkov_etal_PRL65}.

Our measurements on the relaxation as a function of the precession
frequency show that the main source for the temperature-independent
contribution to the relaxation is radiation damping~\cite{raddamp}: The
precessing magnetization of the condensate induces a current in the NMR
pick-up coil which leads to dissipation in the active impedance of
the measuring electric circuit. When the trapping potential is modified,
the condensate spatial extent changes. This affects dissipation both due to spin
diffusion and radiation damping. Our calculations of the combined effect
are in a good agreement with the observed dependence of the relaxation on
the profile of the applied magnetic field.

\section{Experimental setup}

\begin{figure}
\centerline{\includegraphics[width=0.75\textwidth]{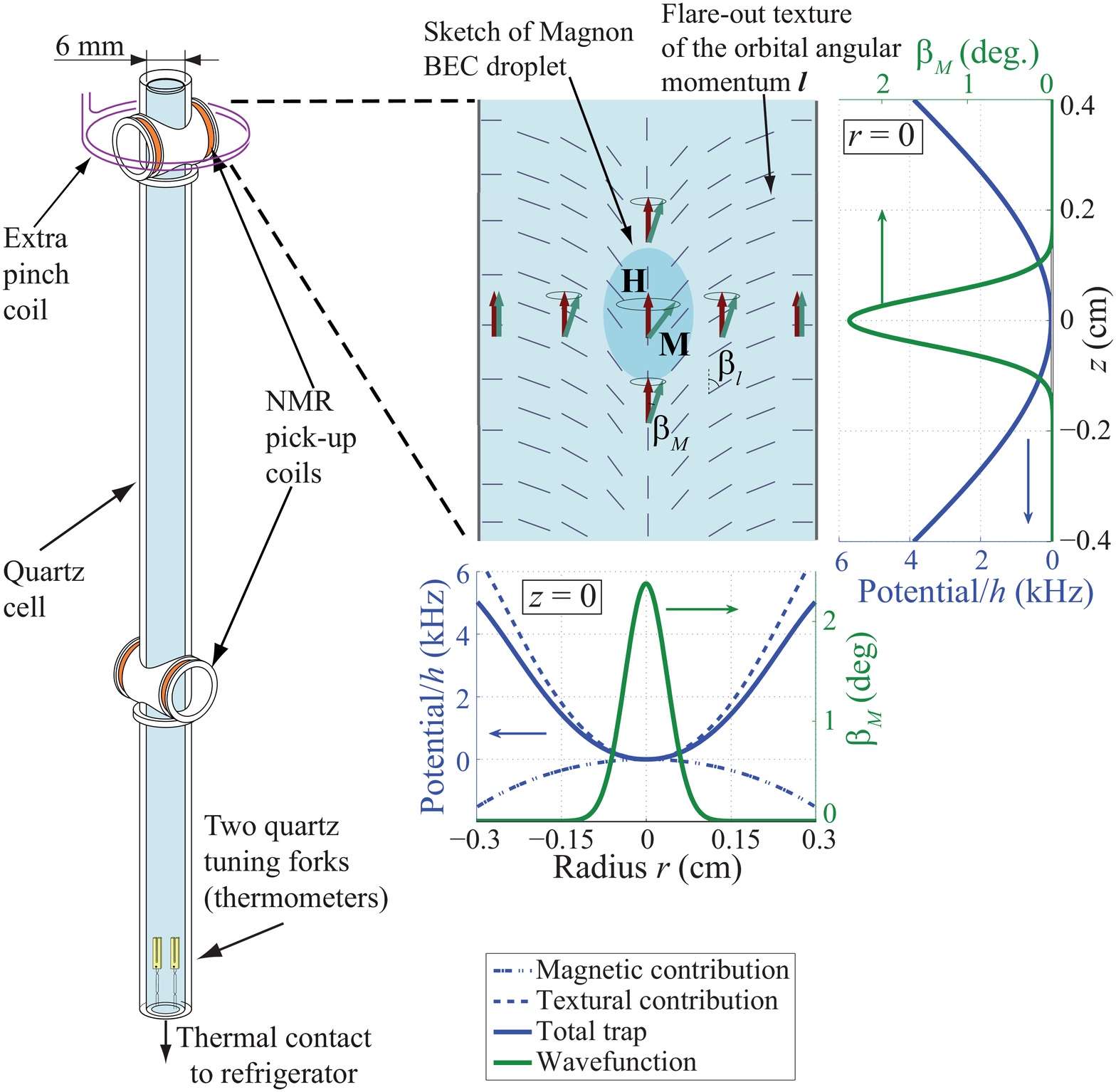}}
\caption{\label{fig:meas_setup}(Color online) Sketch of the experimental
  setup and calculated trapping potentials for
  magnons. \textit{(Left)} The sample
  container is
  equipped with two NMR spectrometers and tuning-fork
  thermometers. In the upper spectrometer, which is used in this
  work, a special pinch coil provides a minimum of the polarizing field in
  the axial direction. \textit{(Right)} A sketch of the magnon condensate. In the
  condensate the magnetization $\mathbf{M}$ is precessing around the
  axially oriented magnetic field $\mathbf{H}$ with the same frequency and
  coherent phase despite magnetic field and textural
  inhomogeneities. The orientation of the orbital anisotropy axis
  $\hat{\mathbf{l}}$ is shown with short segmented lines. The two plots
  show the trapping potential for $P=0.5\,$bar, $T=0.14T_{\mathrm{c}}$, and current $I_{\mathrm{min}} = 2.0\,$A in the pinch coil, calculated respectively at $r=0$ or at $z=0$ (blue lines). The trap is a combination of the axial
  minimum in the Zeeman energy and the interaction energy between the
  precessing spin and the orbital texture which provides trapping in
  the radial direction. The profile of the ground-state wave function
  in this potential is shown with green lines.}
\end{figure}

The superfluid $^3$He-B sample fills a 15\,cm long cylindrical
container made from fused quartz, Fig.~\ref{fig:meas_setup}. The
center of the magnet system used for these measurements is placed at
10\,mm from the upper wall of the cylinder. The static magnetic field
of about 26\,mT for the NMR measurements is oriented along the axis of
the cylinder and has an inhomogeneity $\Delta H/H \approx 4 \cdot 10^{-4}$. An additional pinch
coil creates the field in the opposite direction to that of the main
solenoid and thus the total field has a minimum in the axial
direction. The depth of the minimum can be controlled in the range
0--1\,mT by changing the current $I_{\mathrm{min}}$ in the pinch coil in the
range 0--4\,A. The NMR pick-up coil has two sections wound from
copper wire and is part of a tank circuit tuned to 826\,kHz
frequency with a Q value of about 130. The same coil is used to
pump magnons.

Temperature is measured using a quartz tuning fork thermometer~\cite{fork_JLTP}
installed inside the sample tube approximately 20\,mm above its lower
end, which opens to the liquid $^3$He volume with the sintered heat exchanger on the nuclear cooling stage. As the
scattering of thermal quasiparticles from the sample walls is probably
diffusive and there is a heat leak to the sample of the order of
12\,pW~\cite{hosio_thermal_PRB84}, some temperature difference is expected along the
cylinder from the NMR spectrometer to the thermometer fork. However, this possible temperature difference is small and does not affect the interpretation of results, as discussed below.

The magnetic part of the trapping potential is due to the Zeeman energy
$F_{\mathrm{Z}} =\hbar\omega_{\mathrm{L}}(r,z)|\Psi|^2$, where
$\omega_{\mathrm{L}}(r,z)=\gamma H(r,z)$ is the local Larmor frequency in the
magnetic field $H(r,z)$, $\gamma$ is the absolute value of the gyromagnetic ratio, $\Psi$ is the wave function of the magnon
condensate, and the magnon density is $|\Psi|^2 = (S-S_z)/\hbar = (\chi
H/\gamma\hbar) (1-\cos \beta_M)$. Here
$S=\chi H/\gamma$ is the equilibrium spin density, $\chi$ is the magnetic susceptibility of the B
phase, and the magnon density is characterized in terms of the
tipping angle of magnetization $\beta_M$: $S_z = S
\cos\beta_M$. As shown by the profiles of the potential in Fig.~\ref{fig:meas_setup}, our coil system produces an approximately quadratic minimum of the Zeeman energy in the axial direction and an approximately quadratic maximum in the radial direction.

The textural part of the trapping potential comes
from the spin-orbit interaction energy~\cite{bunvol_review}
\begin{equation}
    \label{eq:spin-orbit}
    F_{\mathrm{so}}=\frac{4\Omega_\mathrm{B}^2}{5\omega_{\mathrm L}}\sin^2\frac{\beta_l (r,z)}{2}|\Psi|^2.
\end{equation}
Here we omitted terms quadratic in magnon density. The strength of the
spin-orbit interaction is characterized by the Leggett frequency
$\Omega_{\mathrm{B}}$. The spatial dependence comes from the variation of the angle $\beta_l$ between the
orbital anisotropy axis $\hat{\mathbf{l}}$ and the magnetic field. In a
cylindrical sample in the axial magnetic field the texture of
$\hat{\mathbf{l}}$ usually takes an axially-symmetric flare-out form, so
that $\beta_l = 0$ at the axis and grows to $\beta_l =\pi/2$ at the
cylindrical wall. According to Eq.~(\ref{eq:spin-orbit}) this results
in a minimum of the spin-orbit energy in the radial direction, which is approximately
quadratic at the bottom, as $\beta_l$ close to the axis changes linearly with distance $r$ from the axis.

The total confinement potential is the sum of the Zeeman and spin-orbit
interaction energies. Close to the center of the trap it turns out to be nearly
harmonic, and thus the spectrum of standing spin waves becomes the
familiar spectrum of eigenstates in an axially symmetric harmonic trap~\cite{QM_intro_Dover}
\begin{equation}
    \label{eq:harm_full}
    \omega_{n_\phi n_r n_z} = \omega_\mathrm{L}(r=0,z=0) + \omega_r
    (|n_\phi | + n_r +1) + \omega_z (n_z + 1/2),
\end{equation}
where $n_\phi$, $n_r$ and $n_z$ are azimuthal, radial and axial quantum numbers, respectively. Their allowed values are:
\begin{equation}
        n_\phi = 0, \pm 1, \pm 2, \ldots, \quad
        n_r = 0, 2, 4, 6, \ldots, \quad
        n_z = 0, 1, 2, \ldots.
\end{equation}
In Eq.~(\ref{eq:harm_full}) $\omega_r$ and $\omega_z$ are the radial and axial oscillator (or
trapping) frequencies, respectively. By measuring the current induced
in the NMR pick-up coil we actually measure the transverse
magnetization of the sample,
$M_\perp = \chi H \int \sin\beta_M(r,z)dV$. Thus we can observe
states only with the azimuthal number $n_\phi=0$ because of the
antisymmetricity of the magnon wave function and $\beta_M$ for
other values of $n_\phi$. For the same reason only states with even axial
number $n_z$ are seen. Thus the spectrum of the lowest-lying magnon
states in our trap is
\begin{equation}
    \label{eq:harm_fit}
    \omega_{n_rn_z} = \omega_\mathrm{L}(r=0,z=0) + \omega_r (n_r+1) + \omega_z (n_z+1/2)
\end{equation}
with even $n_r$ and $n_z$. Note that for these symmetric states
$M_\perp\propto\int |\Psi| dV \propto \sqrt{N_{\mathrm{m}} V_{\mathrm{m}}}$, where $N_{\mathrm{m}}$ is the number of magnons
in the trap and $V_{\mathrm{m}}$ is the volume occupied by the condensate.

\section{Spectroscopy of the magnon levels in the trap}

Pumping of magnons with continuous-wave (cw) or pulsed
NMR techniques is usually coherent, since the frequency of rf pumping,
$\omega_{\mathrm{rf}}$,
fixes the magnon chemical potential, and the orientation of the rf field sets
the phase of the precession of the magnetization. As the magnons in
the trap relax towards the ground state (and lower
precession frequencies), this coherence
is lost, though, and the ground-state condensate develops
\textit{spontaneously} coherent precession at the frequency
$\omega_{00} < \omega_{\mathrm{rf}}$ with the phase of its precession unrelated to
that of the rf excitation field~\cite{cousins_PRL82, magnon_PRL}. This remarkable feature
of Bose-Einstein condensation has also been demonstrated using
incoherent pumping with white noise~\cite{fisher_physicaB}.

Thus, observation of the ground-state
magnon BEC is reasonably straightforward when a proper 3-dimensional
trap is created and the temperature is sufficiently low to increase the lifetime of the condensate. Pumping then has to be applied at a frequency above $\omega_{00}$ either as a spectrally wide pulse or, if using
narrow-band cw-NMR excitation, it has to coincide with one of the
excited levels $\omega_{n_rn_z}$ in the trap. The
response is then measured at $\omega_{00}$ frequency. The
off-resonant excitation is practical since it allows to
separate the frequencies of excitation and measurement and to use the
same NMR coil for both purposes without interference.

\begin{figure}
\centerline{\includegraphics[width=\textwidth]{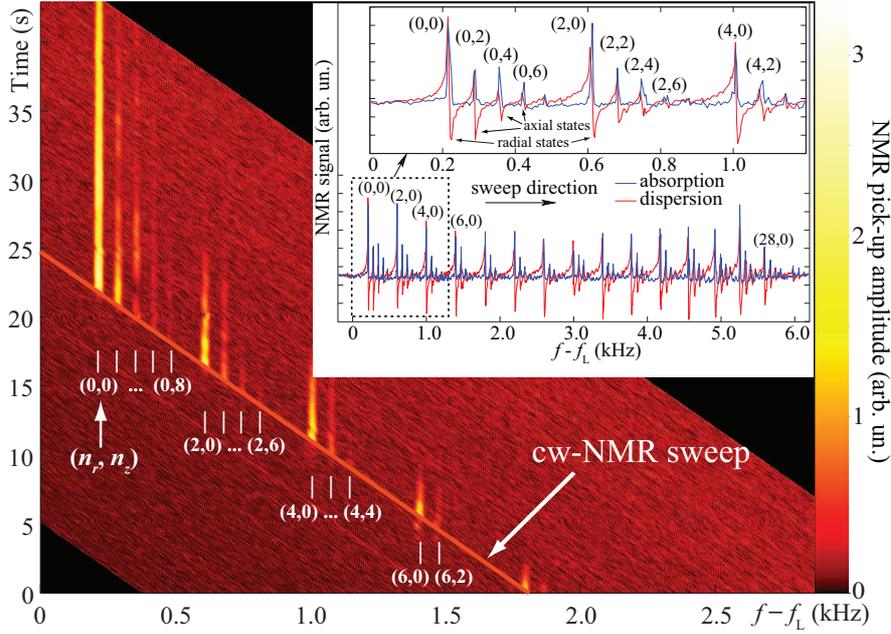}}
    \caption{\label{fig:cw-sweep}(Color online) Measurement of the
  magnon levels in the trapping potential using cw-NMR sweep at a very
  low excitation amplitude. The main plot shows the color-coded amplitude
  of the Fourier spectrum of the signal from the NMR pick-up coil as a
  function of time. The time window of the Fourier transform is
  0.8\,s. The vertical white lines mark the magnon levels $(n_r,n_z)$ in the harmonic
  approximation, Eq.~(\ref{eq:harm_fit}), with $\omega_r / 2\pi =
  197\,{\textrm{Hz}}$, $\omega_z / 2\pi = 35\,{\textrm{Hz}}$ and
  $f_{\mathrm{L}} = \omega_{\mathrm{L}}/2\pi=826\,{\textrm{kHz}}$. The vertical measured
  traces emerging at these frequencies represent coherent precession of
  the magnon condensates at the corresponding trap levels. The longest trace belongs to the ground state
  $n_r=n_z=0$. The \textit{inset} shows the traditional cw-NMR spectrum measured at the same conditions as in the main panel: $P=0.5\,{\textrm{bar}}$, $T=0.14T_{\mathrm{c}}$, and $I_{\mathrm{min}}=0.25\,{\textrm{A}}$. Absorption peaks are seen at the frequencies where magnons are injected to the trap levels with quantum numbers marked near the peaks.
}
\end{figure}

To determine the frequencies $\omega_{n_rn_z}$ of all excited levels accurately, a special approach should be used. We rely on the
experimental observation (which remains so far unexplained) that for the
off-resonant excitation to be effective some excitation threshold
should be overcome. In the experiment we decrease the amplitude of rf
pumping until we notice that magnons pumped to the excited levels do not relax to the ground level. Then the frequency $\omega_{\mathrm{rf}}$ of the cw pumping can be swept to determine the positions of all
levels. An example of such a measurement is presented in Fig.~\ref{fig:cw-sweep}. In the main panel the amplitude of the Fourier spectrum of the signal from the NMR coil is shown as a function of
time. The diagonal line represents the frequency sweep of the NMR
excitation. When it crosses the frequencies of the magnon levels, the corresponding precessing signals are excited as shown by the
decaying vertical traces. In this example, the precession frequencies of the condensates
do not change during the decay since the magnon density is so small from
the start that the self-modification of the textural potential is
negligible.

The magnons on the excited levels of the trap demonstrate spontaneously coherent
precession, characteristic for the BEC: The precession frequency during the decay is not
related to the rf pumping frequency, which is continuously changing,
and the decay time of several seconds significantly exceeds the decoherence
time of linear NMR, which in our conditions is less than 3\,ms. This is
another demonstration of the excited-level BEC discussed in Ref.~\cite{magnon_PRL}.

More standard cw-NMR spectra of standing spin waves can be measured with the detection locked to the excitation frequency, see inset in Fig.~\ref{fig:cw-sweep}. A small enough pumping amplitude and a sweep towards higher frequencies, unlike in Refs.~\cite{magnon_PRL,vortex_core_JLTP162}, are used to avoid off-resonant excitation and a large non-linear signal due to the phenomenon of self-trapping. Thus the frequencies of the maxima in the absorption signal provide a good measure of the magnon levels $\omega_{n_rn_z}$. All the radial states up to the limit put by the Leggett frequency $\Omega_{\mathrm{B}}$ in Eq.~(\ref{eq:spin-orbit}) can be observed. It should be noted that such traditional cw-NMR measurements have limited use in the studies of the trapped magnon condensates since they cannot provide information about the off-resonant excitation or the long lifetime of the different condensate states.

\begin{figure}
\centerline{\includegraphics[width=\textwidth]{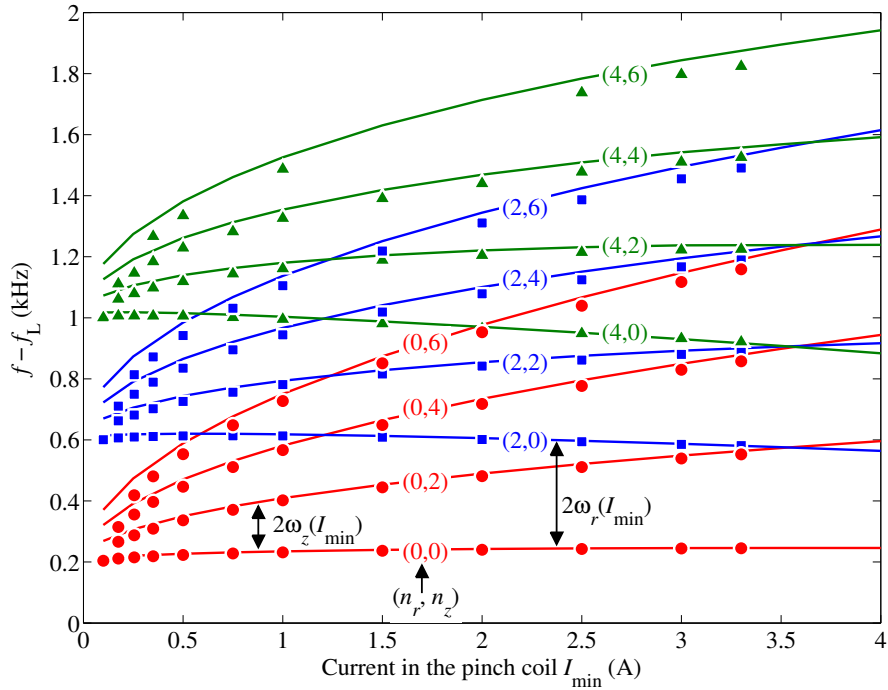}}
    \caption{\label{fig:mod_Imin} (Color online) Frequencies of the
  lowest-lying energy levels in the magnon trap as a function of
  the current $I_{\mathrm{min}}$ in the coil creating the minimum in the
  polarizing field. Measurements (symbols) are compared to
  calculations (lines) of the eigenstates of the model trapping
  potential. The measurements are performed at $T = 0.15T_{\mathrm{c}}$ but in
  practice the trapping potential is temperature-independent below
  $0.2T_{\mathrm{c}}$.}
\end{figure}

A few of the lowest magnon energy levels are plotted as a function of the
current in the minimum coil $I_{\mathrm{min}}$ in
Fig.~\ref{fig:mod_Imin}. These follow reasonably well the
harmonic approximation Eq.~(\ref{eq:harm_fit}) with $\omega_z
\propto \sqrt{I_{\mathrm{min}}}$ and $\omega_r$ decreasing approximately
linearly with $I_{\mathrm{min}}$. These dependences are expected for the
axial minimum of the Zeeman energy with the depth proportional to
$I_{\mathrm{min}}$ and the radial maximum of the Zeeman energy which adds
to a deeper radial minimum of the textural energy, see
Fig.~\ref{fig:meas_setup}. The textural energy itself does not change
substantially due to the small change of the static field provided by
the pinch coil. In most of the experimental range $\omega_z <
\omega_r$, i.e., the condensate droplet is elongated in the axial
direction. As the quantum numbers get larger,
the harmonic approximation breaks, because such states probe the texture with larger $\beta_l$ closer to the wall where $\sin \beta_l$ is no more proportional to $r$, and because of the anharmonicity of the field profile at larger distances from the center of the trap.

In Fig.~\ref{fig:mod_Imin} we compare the measured level frequencies
with the calculations beyond the harmonic approximation. The solid
lines represent the
solution of the eigenvalue problem in the realistic model of the
3D trapping potential. In this model we calculate the textural part of the
potential using the methods of Refs.~\cite{texture_thune,texture_kopu} without fitting parameters. The
magnetic part of the potential is calculated from the geometry of the
solenoidal superconducting coils with an extra fitting parameter for the screening of the
magnetic field by the superconducting parts of the setup. The agreement between calculations and
experiment in Fig.~\ref{fig:mod_Imin} is reasonably good, and the same
model of the magnetic field properly describes also the shape of the NMR response in
normal $^3$He. In the next section we use
the shape of the ground-state wave function in the model potential to
analyze quantitatively the measurements of the relaxation of the
ground-state condensate.

\section{Relaxation measurements}

\begin{figure}
\centerline{\includegraphics[width=\textwidth]{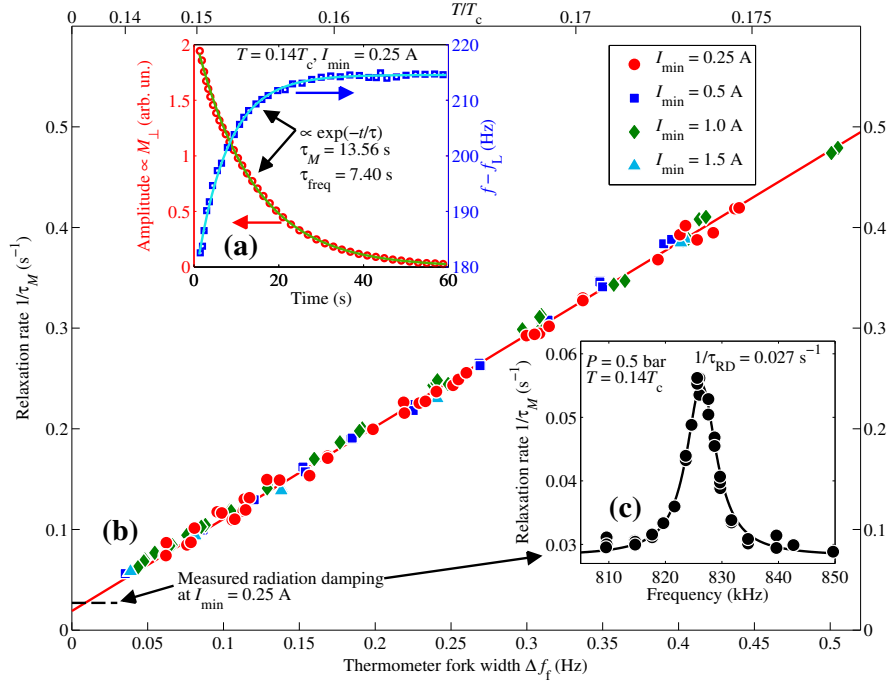}}
\caption{\label{fig:relax_vs_T}(Color online) Relaxation rate of
  the magnon BEC. (a) An example of the relaxation measurement where
  the amplitude (circles) and the frequency (squares) of the precessing signal are fit with the exponential decay curves (lines). (b) Dependence of
  the relaxation rate $1/\tau_M$ of the magnetization on the resonance
  width of the thermometer quartz tuning fork (symbols) demonstrates
  nearly linear dependence (solid line) on the density of thermal
  quasiparticles [$\propto\exp(-\Delta/k_{\mathrm{B}}T)$] to which
  the fork is sensitive. Such temperature dependence is explained by
  spin diffusion, while the extrapolated finite zero-temperature
  relaxation is caused by radiation damping. (c) Dependence of
  the relaxation rate on the precession frequency (circles) shows the
  effect of the  $T$-independent radiation damping at
  $I_{\mathrm{min}}=0.25\,$A. The solid line is a fit to the square of
  the Lorentzian response curve of the NMR pick-up resonance circuit,
  where the height of the peak determines the value of the radiation
  damping and the offset in the wings is due to the spin diffusion at
  the temperature of the measurement.}
\end{figure}

After the magnon condensate is created with cw or pulsed NMR and pumping
is turned off, the condensate decays. When the off-resonant excitation
with sufficiently large pumping amplitude is used, then, unlike in
Fig.~\ref{fig:cw-sweep}, magnons from the excited levels in the trap quickly relax
to the ground state, which becomes the only precessing signal living
for a long time. We measure the frequency $f_{\mathrm{m}}(t) = \omega_{00}/2\pi$ and amplitude
$A_{\mathrm{m}}(t)$ of the signal in the pick-up coil induced by the
precessing magnetization of the decaying condensate using Fourier
transformation. An example of such a measurement is shown in
Fig.~\ref{fig:relax_vs_T}a. The final part of the decay is nicely
exponential: $f_{\mathrm{m}}|_t^{\infty}\propto \exp(-t/\tau_{\mathrm{freq}})$ and $A_{\mathrm{m}}(t) \propto \exp(-t/\tau_M)$. The frequency $f_{\rm m} =
(\omega_{\mathrm{L}} + \omega_r(N_{\mathrm{m}}(t)) + \omega_z/2)/(2\pi)$
increases while the number of magnons in the trap $N_{\mathrm{m}}$
decreases as a result of the modification of the textural trapping
potential by magnons which leads to $d\omega_r/dN_{\mathrm{m}} < 0$~\cite{magnon_PRL}. The amplitude $A_{\mathrm{m}}$ depends both on the transverse magnetization of the condensate $M_\perp$ and its geometrical
shape. However, in the final part of the decay when the textural potential
has its final form $A_{\mathrm{m}} \propto M_\perp \propto \sqrt{N_{\mathrm{m}}}$. For small $N_{\mathrm{m}}$, roughly $f_{\mathrm{m}}(N_{\mathrm{m}})|_{N_{\mathrm{m}}}^0\propto M_\perp^2(N_{\mathrm{m}})$~\cite{magnon_sautti}, and this
explains why in the measurements $\tau_{\mathrm{freq}}$ is roughly half
of $\tau_M$.

The measured temperature dependence of the relaxation rate $1/\tau_M$
is plotted in Fig.~\ref{fig:relax_vs_T}b. The horizontal axis
shows the width of the resonance of the thermometer fork. This is to emphasize
that the relaxation rate is an approximately linear function of the
density of thermal quasiparticles. A similar dependence was observed
earlier in Ref.~\cite{relax_lancaster}, where it was connected to the
temperature dependence of the spin diffusion relaxation
mechanism. The spin diffusion coefficient quickly decreases with
temperature at low temperatures owing to the Leggett-Rice effect~\cite{leggett_rice}. Fig.~\ref{fig:relax_vs_T}b also demonstrates that the extrapolation of relaxation to zero temperature (i.e. close to zero
fork width) results in a finite value. This is also consistent with
Ref.~\cite{relax_lancaster} where an approximately
temperature-independent relaxation mechanism was found, in addition to spin
diffusion. With a good knowledge of the magnon condensate
density profile we can extract values for the diffusion coefficient
from the measurements of the relaxation rate. Before that, however, we
proceed to the identification of the temperature-independent relaxation mechanism.

The main assumption in the following considerations is that the decay
of the condensate is relatively slow, so that the superfluid spin
currents within the condensate always support it in a
quasi-equilibrium state. That is, for a given number of magnons
$N_{\mathrm{m}}$ in the condensate we calculate its wave function and
self-consistent trapping potential regarding the condensate as
stationary, as in Ref.~\cite{magnon_sautti}. This results in well-defined dependences for the total energy of the condensate $E(N_{\mathrm{m}})$ and its transverse magnetization $M_\perp(N_{\mathrm{m}})$. Additionally, since the spin-orbit interaction energy is about four orders of magnitude smaller than the Zeeman energy in our conditions, we assume that $E = \int F_{\mathrm{Z}} dV$.

\textit{Radiation damping.} A measurement of the relaxation rate as
a function of the precession frequency $\omega_{00}$ reveals a
dependence similar to the resonant response of the NMR tank circuit,
Fig.~\ref{fig:relax_vs_T}c. Here the frequency $\omega_{00}$ is
controlled by changing $\omega_{\mathrm{L}}$ with the external magnetic
field. Such frequency dependence is explained by the relaxation
mechanism known as the radiation damping~\cite{raddamp}
\begin{equation}
 	\label{eq:raddamp}
	dE/dt = -V_{\mathrm{s}}^2/(2R),
\end{equation}
where $V_{\mathrm{s}}$ is the induced voltage on the
pick-up coil due to the precessing magnetization and $R$ is the active
impedance of the resonance circuit. Expressing $E$ via $M_\perp$ as
noted above we get from Eq.~(\ref{eq:raddamp}) the relaxation
of magnetization with the rate
\begin{equation}
 	\label{eq:tau_RD}
	\tau_{\mathrm{RD}}^{-1} = \frac{V_{\mathrm{s}}^2}{2 R M_\perp}\frac{dM_\perp}{dE}.
\end{equation}
Since $V_{\mathrm{s}}\propto M_\perp$ and $E \approx \hbar \omega_{\mathrm{L}} N_{\mathrm{m}} \propto M_\perp^2$ the rate is approximately
$M_\perp$-independent, i.e., the relaxation is exponential. It should be noted that the radiation damping increases with the increase of the quality factor $Q\propto 1/R$ of the tank circuit. The magnetic flux through the pick-up coil, and thus the induced voltage $V_{\mathrm{s}}$, depends on the profile of the magnon
density in the condensate, but since the trapping potential is almost
$T$-independent in the studied temperature range the radiation damping
is also temperature-independent.

The frequency dependence of $V_{\mathrm{s}}$ induced in the
tank circuit by a given alternating flux through the coil follows the
standard Lorentzian response. Experimentally the radiation damping
can be determined as a difference in relaxation rate between the maximum and the wings in
Fig.~\ref{fig:relax_vs_T}c. This value is plotted in
Fig.~\ref{fig:relax_vs_T}b by the horizontal dashed line. As one can
see, the radiation damping explains an essential part of the
zero-temperature damping of the magnon BEC. The conclusion whether
actually all the damping is
explained in this way depends on the precise knowledge of the zero-temperature (or intrinsic) width of the thermometer fork, $\Delta f_{\mathrm{f}}^{\mathrm{i}}$. Unfortunately, this
value is not known with sufficient precision, and thus is not taken into account in Fig.~\ref{fig:relax_vs_T}. On the two
cool-downs of the cryostat during the course of these measurements the
width of the fork resonance in vacuum was measured as 12 and 17\,mHz
at $T=12-13\,$mK, while the extrapolation of the $1/\tau_M$ temperature
dependence in Fig.~\ref{fig:relax_vs_T}b reaches $1/\tau_{\mathrm{RD}}$
value at the fork width of $(9\pm 4)$\,mHz. This is consistent with the measured value $\Delta f_{\mathrm{f}}^{\mathrm{i}}=12\,$mHz thus leaving no indications for additional damping mechanisms at zero temperature. If $\Delta f_{\mathrm{f}}^{\mathrm{i}}=17\,$mHz, then the unexplained zero-temperature damping of the magnon BEC would amount to about 25\% of the observed radiation damping.

We might also note that no saturation is seen in the temperature dependence
of relaxation at the lowest temperatures in
Fig.~\ref{fig:relax_vs_T}b. This means that the temperature gradient
along the sample column is sufficiently small and does not result in effects larger than the
scatter in the measurements. The residual temperature difference
between the thermometer and the NMR volume can be considered as a
contribution to the uncertainty in $\Delta f_{\mathrm{f}}^{\mathrm{i}}$.

\begin{figure}
  \includegraphics[width=0.5\textwidth]{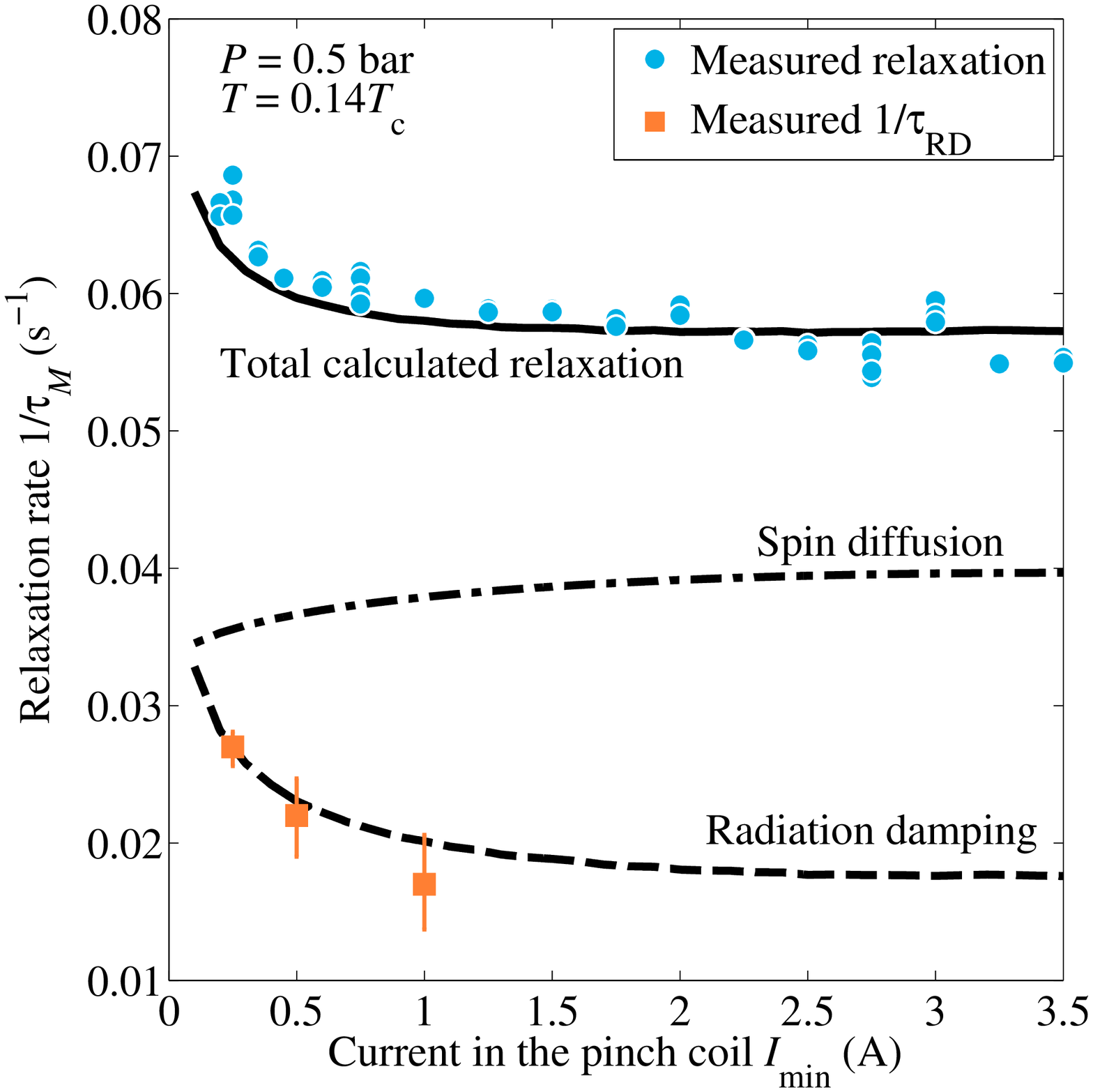}
  \hfill
  \includegraphics[width=0.5\textwidth]{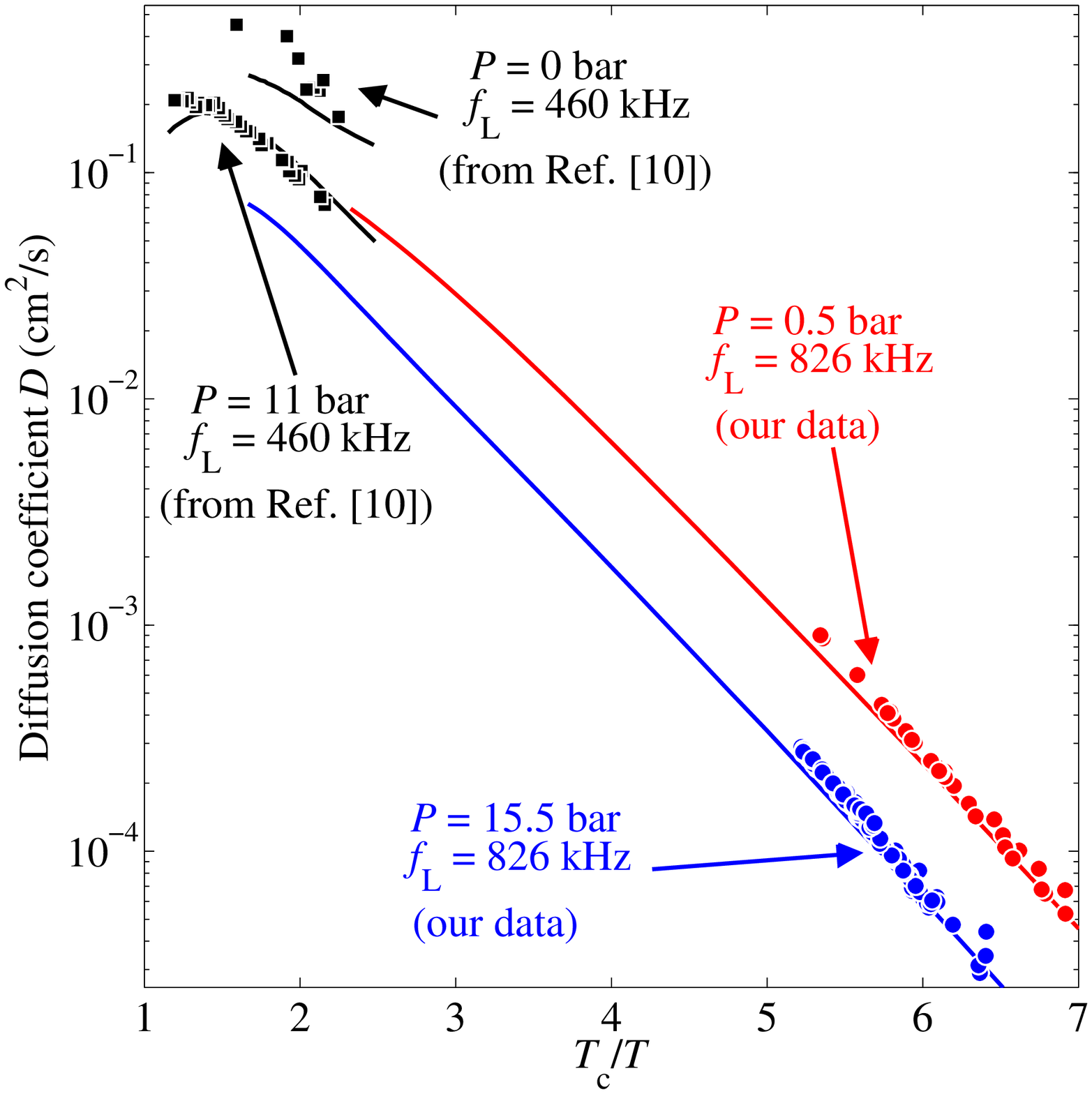}
\caption{\label{fig:relax_vs_Imin}(Color online) \emph{(Left)}
  Relaxation rate of the magnon condensate as a function of the
  current $I_{\mathrm{min}}$ in the field minimum coil
  (circles). Independent measurements of the radiation damping executed in similar way as in
  Fig.~\ref{fig:relax_vs_T}c are shown with squares. The lines show the
  calculated dependencies for the radiation damping (scaled vertically
  to match the experimental values), for the spin diffusion with the
  diffusion coefficient taken from the right panel, and the total
  relaxation. \emph{(Right)} The temperature dependence of the spin
  diffusion coefficient. Values extracted from the measurements in
  Fig.~\ref{fig:relax_vs_T} at $I_{\mathrm{min}} = 0.25\,$A (red circles) and from similar measurements at $P=15.5\,$bar and $I_{\mathrm{min}} = 2\,$A (blue circles) are
  compared to the theoretical prediction~\cite{Einzel_JLTP84} (lines) and to the measurements at higher temperatures
  \cite{Bunkov_etal_PRL65} (squares). Note that the measurements are done at different frequencies while, according to the theory, $D$ is approximately inversely proportional to $f_{\mathrm{L}}$~\cite{Einzel_JLTP84}.}
\end{figure}

\textit{The spin diffusion} contribution to the relaxation of the
condensate is expressed as~\cite{fomin_relax,markelov_spindiff}
\begin{equation}
    \label{eq:zeem_dissip}
    \frac{dE}{dt}=-\frac{D\gamma^2}{\chi}\left\langle\frac{\partial S_i}{\partial x_j}\frac{\partial S_i}{\partial x_j}\right\rangle,
\end{equation}
where $D$ is the transverse component of the spin-diffusion
tensor and indices $i$ and $j$ correspond to
spin components in the cartesian coordinates. The brackets denote
averaging over the volume. The spin components can be expressed via the
condensate wave function, and assuming its phase to be spatially
uniform we arrive to the expression for the relaxation rate of the
magnetization
\begin{equation}
     \label{eq:tau_SD}
    \tau_{\mathrm{SD}}^{-1}=\frac{\omega_{\mathrm L}^2\chi}{\gamma^2}\frac{D}{M_\perp}\frac{dM_\perp}{dE}
    \left[\int(\nabla\beta_M)^2 dV\right].
\end{equation}
We use this expression to fit the relaxation data in
Fig.~\ref{fig:relax_vs_T}b using $D$ as a fitting parameter. We
account only for the spin-diffusion and radiation-damping
contributions to the relaxation, i.e., express $ \tau_M^{-1} =
\tau_{\mathrm{SD}}^{-1} + \tau_{\mathrm{RD}}^{-1}$, and thus self-consistently
take $\Delta f_{\mathrm{f}}^{\mathrm{i}}=9\,$mHz in the temperature calibration. In calculating $\beta_M(r,z)$ and $dE/dM_\perp$ in
Eq.~(\ref{eq:tau_SD}) we use self-consistent calculations of the
textural trapping potential and the wave functions. The texture modification by the
condensate is relatively small, and the calculated relaxation rate is approximately independent of $M_\perp$. This is in agreement with the experiments.

The results for the temperature dependence of the diffusion coefficient are shown in the
right panel of Fig.~\ref{fig:relax_vs_Imin}. The expected
nearly-exponential decrease of $D$ at the lowest temperatures is
observed. The measurements are compared to the theoretical prediction
from Eq.~(108) in Ref.~\cite{Einzel_JLTP84}. In evaluating this expression we take the Fermi liquid parameters from
Ref.~\cite{Dobbs_page_52}, and we use the weak-coupling plus
gap~\cite{wcp-gap}. As in Ref.~\cite{Einzel_JLTP84}, we approximate the
order-parameter $\hat{\mathbf{n}}$ vector to be oriented uniformly along the
field direction. In principle this is not strictly valid for the trapped
magnon BEC, but in Eq.~(\ref{eq:zeem_dissip}) the main contribution
comes from the condensate region where the deflection angle of
$\hat{\mathbf{n}}$ from the magnetic field is below $15^\circ$, so we
consider this assumption to be reasonable. Overall, we consider the agreement
between the measurements and the theoretical value of $D$ as good,
given that the theoretical expression has no fitting parameters.

\textit{Dependence of relaxation on the trapping potential.} Both the
spin diffusion and the radiation damping depend on the
magnon condensate wave function. When the axial confinement of the
condensate increases with increasing $I_{\mathrm{min}}$, spin diffusion
increases due to the increase in the gradients of the spin. Simultaneously,
radiation damping decreases due to the reductions in the spatial extent
of the condensate and the effective filling factor for the
NMR pick-up. The combined effect is measured in the left panel of
Fig.~\ref{fig:relax_vs_Imin}. It is in a reasonable agreement with our
calculations of the respective changes in the damping.

\section{Conclusions}
We have studied the relaxation of trapped Bose-Einstein
condensates of mag\-nons in superfluid $^3$He-B at temperatures below
$0.2T_{\mathrm{c}}$. For the first time the tem\-pe\-ra\-ture-in\-de\-pen\-dent relaxation
mechanism from radiation damping
is identified. The measurements of the relaxation from
spin diffusion allow us to determine reliably
the spin diffusion coefficient $D$ at the lowest temperatures. The measured
values of $D$ agree well with the theoretical prediction and the earlier
measurements at higher temperatures. We have demonstrated the dependence
of the relaxation on the shape of the trapping potential which is
important for a proper analysis of the relaxation in the regime of large magnon
density, where the self-modification of the trapping potential results in non-exponential effects. The achieved understanding of
the relaxation of the magnon condensates in bulk $^3$He-B provides a firm
basis for studying relaxation effects originating from exotic fermion
bound states at the surfaces of $^3$He-B~\cite{majorana_surface} or in the cores of quantized vortices~\cite{majorana_vortex}.

\begin{acknowledgements}
We thank Yu.M. Bunkov, V.V. Dmitriev, P. Hunger, P. Skyba and G.E. Volovik for useful discussions.
This work has been supported in part by the EU 7th Framework Programme (FP7/2007-2013, Grant No. 228464 Microkelvin) and by the Academy of Finland through its LTQ CoE grant (project no. 250280). P.J.H. and J.J.H. acknowledge financial support from the V\"{a}is\"{a}l\"{a} Foundation of the Finnish Academy of Science and Letters.
\end{acknowledgements}

%%%%%%%%%%%%%%% BibTeX users please use one of
%\bibliographystyle{apsrev4-1}% For RevTex users
%\bibliographystyle{spphys}% For APS-like style for physics
%\bibliographystyle{spmpsci}
%%%%%% For umlaut use $\ddot{\mathrm{o}}$ not\''{o}%%%%%%%%%%%%
%%%%  This is a well known problem for BibTex and latex2e%%%%%%
%%%%%%%%%%%%%% AND uncomment both lines below.
%\bibliography{references_magnon}   % Name.bib is author's BibTeX data base
%\end{document}
% Non-BibTeX users please use

\newpage

\bibliographystyle{plainnat}

\end{document}